\newif\ifsubmission
\submissiontrue

\documentclass[10pt,conference]{IEEEtran}

\usepackage{cite}
\usepackage[autolanguage]{numprint}
\npdecimalsign{.}
\usepackage[binary-units=true, per-mode=symbol, per-symbol=/, range-phrase=-, range-units=single, detect-all, group-minimum-digits=3]{siunitx}
\DeclareSIUnit{\M}{\text{M}}
\DeclareSIUnit{\k}{\text{k}}
\usepackage{amsmath,amssymb,amsfonts}
\usepackage{algorithmic}
\usepackage{graphicx}
\usepackage{textcomp}
\usepackage{xcolor}
\def\BibTeX{{\rm B\kern-.05em{\sc i\kern-.025em b}\kern-.08em
    T\kern-.1667em\lower.7ex\hbox{E}\kern-.125emX}}
\usepackage{subcaption}
\usepackage[export]{adjustbox}
\usepackage[bookmarks=false]{hyperref}
\usepackage{cleveref}
\usepackage{xspace}
\usepackage{booktabs}
\usepackage{tikz}
\usepackage{csquotes}
\usepackage{balance}

\expandafter\def\expandafter\UrlBreaks\expandafter{\UrlBreaks
	\do\a\do\b\do\c\do\d\do\e\do\f\do\g\do\h\do\i\do\j%
	\do\k\do\l\do\m\do\n\do\o\do\p\do\q\do\r\do\s\do\t%
	\do\u\do\v\do\w\do\x\do\y\do\z\do\A\do\B\do\C\do\D%
	\do\E\do\F\do\G\do\H\do\I\do\J\do\K\do\L\do\M\do\N%
	\do\O\do\P\do\Q\do\R\do\S\do\T\do\U\do\V\do\W\do\X%
	\do\Y\do\Z}

\newcommand*\circled[1]{\tikz[baseline=(char.base)]{
		\node[shape=circle,draw,inner sep=.2pt] (char) {#1};}}

\ifsubmission
  \newcommand{\TODO}[1]{}
  
\else 
  \newcommand{\TODO}[1]{\textcolor{red}{TODO: #1}}
  
\fi
\newcommand{\coolname}{O\kern.2mm M\kern.2mm G\xspace}
\newcommand{\sanctuary}{SANCTUARY\xspace}

\let\oldS\S
\let\orgautoref\autoref
\renewcommand{\autoref}
        {\def\equationautorefname{Equation}%
         \def\figureautorefname{Fig.}%
         \def\subfigureautorefname{Figure}%
         \def\Itemautorefname{Item}%
         \def\tableautorefname{Table}%
         \def\algorithmautorefname{Figure}%
         \def\paragraphautorefname{Paragraph}%
         \def\sectionautorefname{\oldS}%
         \def\subsectionautorefname{\oldS}%
         \def\subsubsectionautorefname{\oldS}%
         \def\chapterautorefname{Chapter}%
         \def\partautorefname{Part}%
         \def\goalautorefname{Goal}%
         \def\reqautorefname{Req.}%
         \def\adviceautorefname{Rule}%
         \def\parameterautorefname{Param.}%
         \def\definitionautorefname{Definition}%
         \def\theoremautorefname{Theorem}%
         \orgautoref}

\begin{document}
\title{\textsc{Offline} \textsc{Model} \textsc{Guard}:\\ Secure and Private ML on Mobile Devices}

\author{
	\IEEEauthorblockN{Sebastian P.~Bayerl\IEEEauthorrefmark{1}, Tommaso Frassetto\IEEEauthorrefmark{2}, Patrick Jauernig\IEEEauthorrefmark{2}, Korbinian Riedhammer\IEEEauthorrefmark{1}, Ahmad-Reza Sadeghi\IEEEauthorrefmark{2},\\Thomas Schneider\IEEEauthorrefmark{2}, Emmanuel Stapf\IEEEauthorrefmark{2}, Christian Weinert\IEEEauthorrefmark{2}}
	\IEEEauthorblockA{\IEEEauthorrefmark{1}\textit{Technische Hochschule N{\"u}rnberg}, Germany, 
		\{sebastian.bayerl, korbinian.riedhammer\}@th-nuernberg.de}
	\IEEEauthorblockA{\IEEEauthorrefmark{2}\textit{Technische Universit{\"a}t Darmstadt}, Germany, \{tommaso.frassetto, patrick.jauernig, ahmad.sadeghi,\\emmanuel.stapf\}@trust.tu-darmstadt.de, \{schneider, weinert\}@encrypto.cs.tu-darmstadt.de} 
}

\maketitle

\begin{abstract}
Performing machine learning tasks in mobile applications yields a challenging conflict of interest: highly sensitive client information~(e.g., speech data) should remain private while also the intellectual property of service providers~(e.g., model parameters) must be protected.
Cryptographic techniques offer secure solutions for this, but have an unacceptable overhead and moreover require frequent network interaction.

In this work, we design a practically efficient hardware-based solution.
Specifically, we build \textsc{Offline} \textsc{Model} \textsc{Guard}~(\coolname) to enable privacy-preserving machine learning on the predominant mobile computing platform~ARM---even in offline scenarios.
By leveraging a~trusted execution environment for strict hardware-enforced isolation from other system components,~\coolname guarantees privacy of client data, secrecy of provided models, and integrity of processing algorithms.
Our prototype implementation on an~ARM~HiKey~960 development board performs privacy-preserving keyword recognition using~TensorFlow~Lite for Microcontrollers in real time.
\end{abstract}


\section{Introduction}
\label{sec:introduction}

An increasing number of applications running on mobile devices like smartphones and tablets relies on machine learning~(ML) services to enhance the user experience, e.g., to give an estimate on battery life based on user behavior, improve image quality, or perform speech recognition.

Many of these~ML services require frequent cloud inter\-action, resulting in severe privacy risks for billions of users due to the highly sensitive nature of such remotely processed data.
Besides potentially confidential and intimate content, voice recordings, for example, contain unique biometric information that can be abused, e.g., for impersonation attacks and distributing fake recordings.

Privacy breaches in this domain are not fiction: in~2018, a customer requested his recording archive from~Amazon, but accidentally got access to~\numprint{1700} audio files from a stranger~\cite{WashingtonPost}.
Furthermore, state authorities ordered~Amazon to hand out recordings as they might contain evidence of crime~\cite{Independent}.
Media reports also revealed that~Apple, among others, sent voice recordings to third party companies in order to improve their service with manual transcriptions.
The employees of those companies got to listen to private discussions between doctors and patients, business deals, criminal dealings, and sexual encounters~\cite{Guardian_Apple}.
Moreover, biometric data used for identification was recently leaked at a large scale: the database of a~UK government contractor with more than a million fingerprints and facial recognition information was publicly accessible~\cite{Guardian_Breach}.

When relying on online services for mobile~ML applications, there are also usability issues to consider: high latency and, therefore, a bad user experience occurs if the user has an unreliable or low-bandwidth network connection, and high roaming fees may apply if the user is abroad.

A trivial solution for all these issues is to process all sensitive user data on the client's device.
Previously, this approach was severely limited by the storage space constraints on mobile devices and the storage space requirements of~ML models used in practice.
Recently, though, Google lifted this limitation by training a recurrent neural network~(RNN) model for character-level speech recognition and compressing it to only~\SI{80}{\mega\byte}, while delivering the same accuracy as former cloud-based production models with a size of multiple gigabytes~\cite{GoogleAIPaper,GoogleAIBlog}.

However, deploying such a model in unencrypted form is often not in the interest of the service provider.
A production-level model constitutes intellectual property as the underlying training data is usually hard to obtain and creating an accurate while compact model requires extensive expertise~\cite{CNN_IP}.
Furthermore, if attackers have unrestricted model access, the privacy of people represented in the training data is even more threatened by, e.g., membership inference attacks~\cite{MembershipInference} and unintended memorization~\cite{Memorization}.

Cryptographic techniques like homomorphic encryption~(HE) and secure multi-party computation~(SMPC) provide solutions for this conflict of interest:
with~HE, private inputs can be securely processed under encryption by the client or the service provider, whereas with~SMPC, client and server can jointly compute any function on private inputs in a provably secure protocol.
Unfortunately, the computational overhead for~HE when performing complex~ML tasks is impractical for the given mobile scenario, whereas the amount and the frequency of required network communication is the bottleneck for~SMPC protocols.
Thus, we explore hardware-assisted solutions to deliver secure and private~ML on mobile devices in offline scenarios while providing practical efficiency.

{\bf Our Contributions.}
In this work, we build~\textsc{Offline} \textsc{Model} \textsc{Guard}~(\coolname), a generic architecture that efficiently protects machine learning tasks on mobile devices like smartphones and tablets, and demonstrate its practicality using offline keyword recognition as an example application.

\coolname leverages unprivileged~(normal-world) user-space enclaves on~ARM platforms to execute~ML tasks in a hardware-protected environment that is two-way isolated from all other system components to minimize the attack surface.
Utilizing~TrustZone functionality,~\coolname can securely access peripherals like the microphone to protect sensitive information directly from the source.
As a result,~\coolname guarantees complete privacy of client data, secrecy of the provided~ML models, and integrity of processing algorithms.

We provide a fully functional prototype implementation of~\coolname on an~ARM~HiKey~960 development board for offline keyword recognition based on~TensorFlow Lite for Microcontrollers~\cite{TensorFlowLite}.
As~TrustZone on~ARM does not provide user-space enclaves, we leverage~\sanctuary~\cite{SANCTUARY} for our implementation.
Our performance evaluation demonstrates that secure and private offline speech processing is possible in real time even with strong protection guarantees.
As we developed our prototype with~TensorFlow compatibility in mind, our implementation can easily be extended to network architectures used for other related tasks such as end-to-end continuous speech recognition, speaker verification, and emotion recognition.

\section{Related Work}
\label{sec:related_work}

In the following, we review existing works that preserve privacy in machine learning.
The goal there is usually to train a model on the server side without allowing the server to see training data in the clear, or to obliviously classify input data without leaking the model~(inference).
Proposed solutions either rely entirely on cryptography or build on~TEEs.

For protecting only the~IP of~ML models there also exist orthogonal works for model watermarking~\cite{DL_Watermarking} and fingerprinting~\cite{DL_Fingerprinting} that do not consider the privacy of client inputs.

\subsection{Cryptography}
The cryptographic techniques used for privacy-preserving machine learning are~homomorphic encryption~(HE)~and secure multi-party computation~(SMPC).
Also, combinations of these techniques are being studied.
HE allows to perform operations directly on encrypted data, but generally incurs a high computational overhead.
SMPC allows multiple parties to jointly perform secure computations on shared data.
This works by obliviously evaluating a Boolean or arithmetic circuit representation of the desired functionality, but results in a high communication overhead and for some protocols requires interaction for each layer of the circuit.

For cryptographic protocols it is possible to formally prove security with respect to input privacy.
However, many protocols and corresponding implementations assume that both client and server honestly follow the protocol description.
This assumption is unrealistic in real-world scenarios since mobile clients might run modified applications.
Securing such protocols against malicious parties comes at additional cost.

Privacy-preserving neural network inference via~HE and~SMPC was studied in~\cite{orlandi2007oblivious, ICISC09, TIFS11}.
Thereafter, many frameworks for privacy-preserving machine learning have been developed, e.g., \cite{CryptoNets, SecureML, MiniONN, Chameleon, GAZELLE, XONN}.
They allow at least for secure deep/convolutional neural network inference and are usually benchmarked with standard image classification tasks.

Using such cryptographic frameworks requires expert knowledge and thus they are hardly accessible for~ML experts.
However, recently there are efforts to integrate cryptographic protocols into standard~ML tools: for~TensorFlow there are HE~\cite{sealion} and~SMPC~\cite{dahl_private_2018} implementations, and for~Intel's~ngraph compiler there exists HE support~\cite{ngraph-he}.

Unfortunately, the current performance results discourage from actual deployment and scaling them to more involved speech processing tasks seems unrealistic~\cite{Pathak}.
Addressing all outlined disadvantages, with~\coolname we propose a computation- and communication-efficient hardware-assisted design for secure and private~ML on mobile devices that enforces correct execution of the algorithms and can easily be used by~ML~experts due to~TensorFlow~Lite compatibility.

\subsection{Trusted Execution Environments~(TEEs)}
Compared to cryptographic techniques, trusted execution environment~(TEE) architectures provide several orders of magnitude better performance for protecting~ML services~\cite{Slalom}. 
Most of the existing works rely on~Intel~SGX as the dedicated~TEE architecture to protect~ML services.

Ohrimenko et al.~\cite{OSF+16} protect~ML algorithms and models in~SGX enclaves.
They consider a scenario where sensitive data from multiple data providers is aggregated on a remote server while~SGX enclaves are used to protect the training process.
However, the enclaves might leak information to the untrusted software on the server through data-dependent access patterns, which can be exploited in controlled-channel attacks~\cite{liu2015last, xu2015controlled}.
Therefore, the authors develop data-oblivious variants of standard~ML techniques,~e.g., support vector machines, neural networks, and decision trees, which guarantee that all memory accesses do not depend on secret data. 

In Chiron~\cite{Chiron}, an~ML-as-a-Service~(MLaaS) scenario is considered where sensitive data is collected from customers and used for training without revealing the data to the~MLaaS provider. 
This is achieved by performing the training process in a~Ryoan~\cite{ryoan} sandbox (based on~SGX), which protects sensitive customer data but still offers the service provider the possibility to freely select, configure, and train the models.

Myelin~\cite{Myelin} provides security guarantees similar to~\cite{OSF+16} as it relies on data-oblivious deep learning algorithms: every model owner compiles its deep learning model into a privacy-preserving model graph, which is then trained on a remote server~(inside an~SGX enclave) on sensitive data.

In~\cite{CKL+17}, the authors introduce an alternative protection mechanism against controlled-channel attacks that is more efficient and suitable for real-time data processing.
The authors propose to add noise to memory traces by accessing dummy data instead of enforcing data-oblivious memory accesses. 

VoiceGuard~\cite{VoiceGuard} targets the use case of privacy-preserving speech processing.
For this, sensitive voice recordings are collected from user devices, e.g.,~smart home devices like~Amazon~Echo, Google~Home, and~Apple~HomePod, and are sent via secure channels to a service provider.
The service provider performs speech recognition using proprietary models provided by~ML specialists in an~SGX enclave, thereby protecting the user data as well the proprietary models.
The inference results are then securely sent back to the user device.
Very recent work~\cite{ahmed2019prch} also enables efficient private online speech recognition but uses obfuscation techniques and the notion of differential privacy, which significantly degrades accuracy.

In contrast,~MLCapsule~\cite{MLCapsule} considers an offline~MLaaS scenario where the trained model is used on the client side for inference while being protected using an~SGX enclave.

None of the previous works considers the challenge of how user data can be securely collected on the user device.
Intel~SGX, which is mostly used as the dedicated~TEE architecture, is not able to provide a secure communication channel from enclaves to system peripherals, e.g., the microphone or camera~\cite{costan2016intel}.
Thus, sensitive user data is endangered as it could be exfiltrated by malicious software running on the client device.
With~OMG, we present the first~TEE architecture that provides protection for proprietary~ML models and privacy-sensitive user input at the same time.
Furthermore, while~Intel~SGX is a~TEE widely available in recent~Intel~CPUs, most mobile devices like smartphones and tablets come with~CPUs based on the~ARM platform.
This prevents using the previously proposed~SGX-based solutions for securing relevant use cases on mobile devices, e.g., offline speech recognition.
Thus, in this work, we present~\coolname for ARM-based devices and as an example application demonstrate privacy-preserving offline keyword recognition in real time.

\section{Background}
\label{sec:preliminaries}
In the following, we introduce relevant details regarding the~ARM~TrustZone~TEE implementation and the~SANCTUARY security architecture~\cite{SANCTUARY} for user-space enclaves.

\subsection{ARM TrustZone}
\label{subsec:trustzone}

\begin{figure}[!h]
	\centering
	\includegraphics[trim=6mm 51mm 15mm 0cm, clip, scale=0.5]{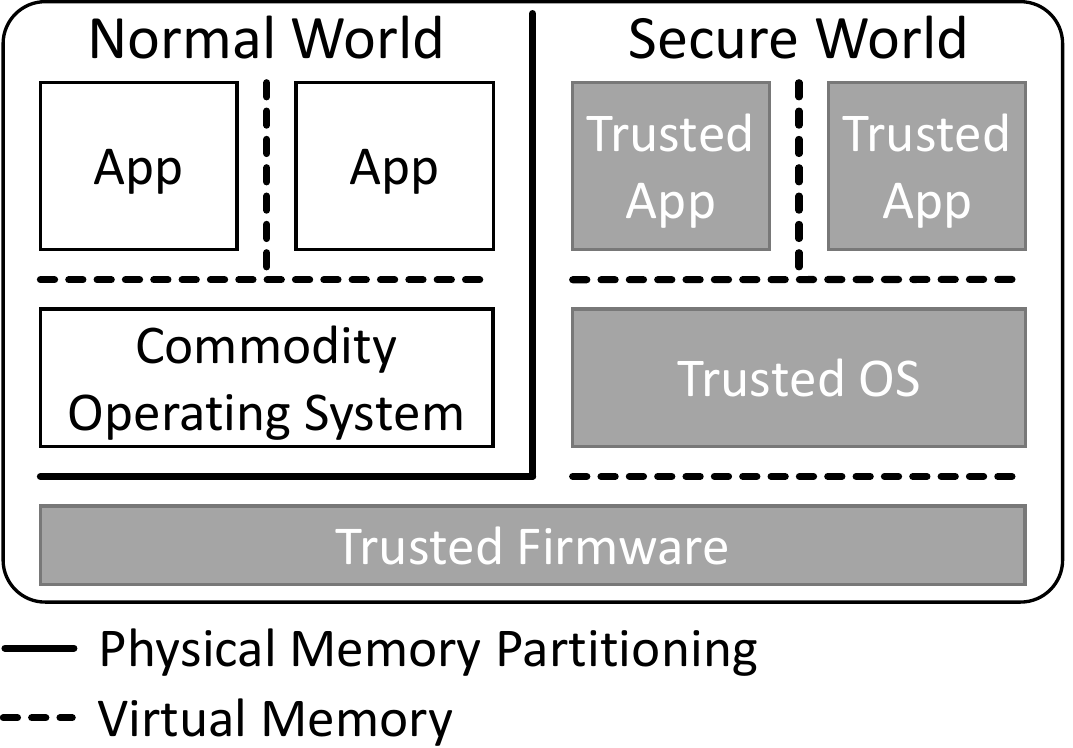}
	\caption{ARM TrustZone architecture overview.}
	\label{fig:tzoverview}
\end{figure}

Trusted execution environments~(TEEs) combine memory isolation techniques~\cite{IMIX,Timberv,Readactor} and attestation~\cite{Trustvisor} with isolated execution to provide protected execution of security-critical code.
For mobile devices, the predominant computing platform is~ARM, which provides a~TEE implementation called~ARM~TrustZone~\cite{TrustZone}.
A chip with~TrustZone capabilities simultaneously runs two security contexts~(or~\enquote{worlds}) as virtual processors: a~\enquote{normal world} and an isolated~\enquote{secure world}~(cf.~\autoref{fig:tzoverview}).
While the normal world executes a commodity~OS~(e.g., Android) and ordinary applications, the secure world forms a~TEE for running security-critical code on a trusted~OS.

A major assumption of~TrustZone is that an attacker cannot compromise code running in the secure world.
Unfortunately, the~TrustZone design is flawed in this aspect: the isolation between applications in the secure world is rather weak and the attack surface is massively increased the more applications run therein~\cite{ProjectZero}.
Thus, the secure world with its privileged platform access is an attractive target for adversaries.

\subsection{\sanctuary}
\label{subsec:sanctuary}

\sanctuary~\cite{SANCTUARY} is a security architecture that circumvents the previously explained flaws of~ARM~TrustZone without requiring hardware extensions, heavy modifications of existing code bases, or major changes in the commodity~OS.
In particular, it allows to run security-critical code in user-space enclaves or so-called~\sanctuary~Apps~(SAs).
SAs are executed in a normal-world environment that is protected via strict hardware-enforced two-way isolation from all other system components to minimize the attack surface.
This is achieved by leveraging~TrustZone's address space controller~(TZASC) to exclusively bind memory to a (temporarily) dedicated~CPU core running an~SA.

\noindent\hspace*{1.35em}The life cycle when running an~SA is as follows:

\begin{enumerate}[
		\setlength{\IEEElabelindent}{0pt}
	]
	\item 
	Setup:~Memory for the~SA instance is prepared by loading the~\sanctuary library~(SL), which is implemented using the~Zircon microkernel~\cite{Zircon}, and the~SA.
	The~TZASC is securely configured to isolate this memory region and the least busy~CPU core is shut down.
	Besides the isolated memory, additional memory regions are shared with the commodity~OS and the secure world, which allows the~SA to access the secure world and~(untrusted)~OS services.
	
	\item Boot: The memory is attested and the CPU core is booted with the~SL providing a basic execution environment.
	
	\item Execution: The~SA runs as a normal-world user process, potentially using services provided by the commodity~OS or secure world code.
	
	\item Teardown: The~CPU core is shut down, data in the first level cache~(L1) is invalidated, the~SA memory is cleaned and unlocked, and finally the~CPU core is handed back to the commodity~OS.
\end{enumerate}

\sanctuary provides code and data integrity as well as data confidentiality, is secure against malicious~SAs, and has no negative impact on the user experience due to the wide availability of multicore chips for mobile devices.
Furthermore, side-channel attacks that extract secrets from caches can be prevented easily since the~L1 cache is core exclusive and the shared second level cache~(L2) can be excluded from~\sanctuary memory without severe performance impact~\cite{SANCTUARY}.

\sanctuary extends~TrustZone to provide an arbitrary number of user-space enclaves.
Additionally, \sanctuary inherits many useful features from~TrustZone like secure boot or~DMA attack protection.
Moreover,~TrustZone allows to assign sensitive peripherals exclusively to the secure world.
An~SA can use this feature by sending communication requests to the secure world code.
After checking the permission rights of the~SA, the secure world reads from the sensitive data and directly stores it in the memory region shared with the~SA.
Thus, performance overhead is only produced by the additional world switches between the~SA and the secure world.

\section{Security Model and Assumptions}
\label{sec:adversary}

In this paper, we consider two parties collaborating to perform~ML tasks on sensitive data provided by one party while protecting the intellectual property of the other party.

The~\emph{user U} provides input data to be processed.
She is concerned about the privacy of the content to be processed~(i.e., her inputs as well as outputs) and biometric characteristics potentially used throughout processing.
Lastly, the user does not want to be traceable across multiple sessions.

The~\emph{vendor V}~(who might act as the service provider) provides~ML algorithms including corresponding models.
The models constitute the vendor's intellectual property, hence the user must not be able to reverse engineer, share, or break the license check of these models.

{\bf Adversary Model.}
The adversary's goal is to extract sensitive information, i.e., the intellectual property of the vendor, the input and output of the user, or data that allows the adversary to identify or track the user.
We assume that the adversary is in control of the user's device.
The adversary has full control over the software running in the normal world of the user's device, including privileged software like the commodity~OS.
We assume that the adversary cannot perform hardware attacks, e.g., a physical side channel to extract secret keys.
For the enclave we assume that all of~\sanctuary's defense mechanisms are in place, including hardware cache partitioning (for a detailed discussion see~\cite{SANCTUARY}).
\section[OMG Design]{\coolname Design}
\label{sec:solution}

\begin{figure}[t]
	\centering
	\setlength{\fboxsep}{0pt}
	\fbox{\includegraphics[trim=0cm 46.3mm 65.9mm 0cm, clip, width=\columnwidth]{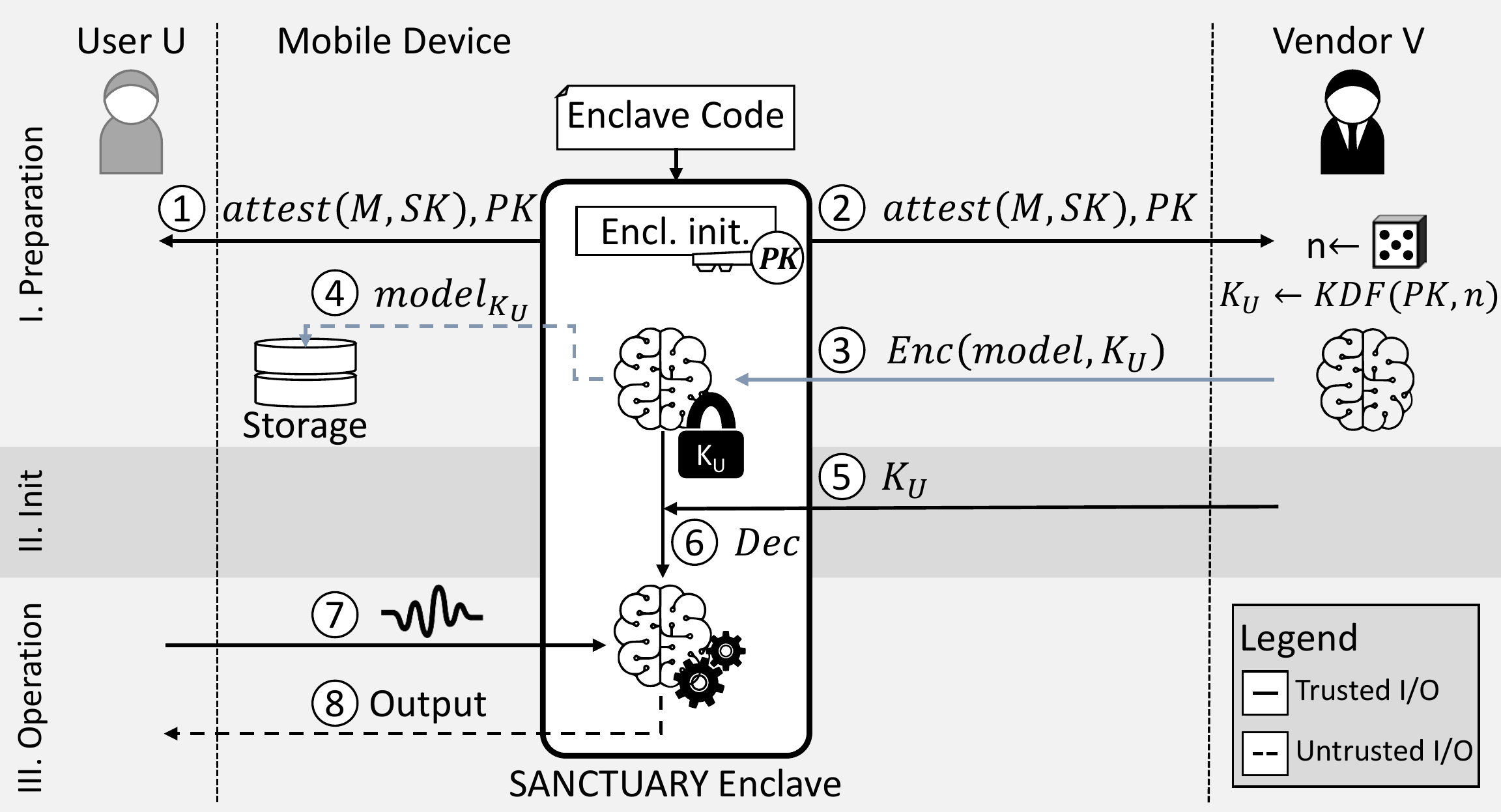}}
	\caption{\coolname overview. Once the encrypted model is stored locally, steps in gray are optional until a model update.}
	\label{fig:arch}
\end{figure}

\coolname enables privacy-preserving and efficient offline execution of~ML algorithms on untrusted~ARM-based systems.
For the sake of simplicity, we explain our solution based on the speech recognition scenario visualized in~\autoref{fig:arch}.

The vendor~V's private input consists of a~ML model.
The user~U's private input consists of voice recordings.
In this example, the~ML model is the vendor's intellectual property and any information about its architecture or trained weights must never be disclosed.
The only output is the transcription, which is sent to the user.

\coolname works in three phases:~(I.)~preparation,~(II.)~initialization, and~(III.)~operation.
In the preparation phase, the enclave~(containing the~SL and~SA) is loaded and attested to user~U and vendor~V. Then,~V provides the encrypted~ML model to the enclave.
In the initialization phase,~V sends the decryption key for the~ML model so that the enclave can decrypt the model.
Finally, in the operation phase, the enclave is ready to perform offline speech recognition.
U sends her voice recordings to the enclave and receives respective textual output~(which can be further processed into an action, as with virtual assistants).
Next, we detail the individual phases:

{\bf I.\ Preparation Phase.}
First, the enclave needs to be run on~U's device.
The enclave contains the environment required to apply the~ML model to input data.
The enclave code can be open source, since it does not contain any vendor secrets~(e.g., it may just consist of a~TensorFlow environment), and can be distributed by the device manufacturer via regular distribution channels.
To load the enclave, its code is first copied to memory and locked to a dedicated~\sanctuary~CPU core so it cannot be changed anymore by the commodity~OS~(cf.~\autoref{subsec:sanctuary}).
Then, the enclave is attested~(\enquote{measured}) by~\sanctuary, i.e., a cryptographic hash of the initial memory content of the enclave is created and stored securely.
If the enclave code is manipulated before the creation process, the measurement will produce a different result and the manipulation will be detected.

\sanctuary then assigns an unique asymmetric key pair to this enclave, e.g., by using RSA~\cite{RSA}~(the public key~$\textit{PK}$ is shown in~\autoref{fig:arch}).
This key pair is derived from the platform certificate issued by the device vendor, effectively creating a certificate hierarchy similar to~SSL certificates.
To assure to~U that the correct enclave code has been loaded, an attestation report is generated~(i.e., the cryptographic hash of the initial memory content is signed using the secret key~$\textit{SK}$ corresponding to~$\textit{PK}$) and sent to~U using the secure output functionality of \sanctuary~\circled{1}.
Such an attestation report is also sent to~V using a secure connection~(e.g., via TLS) directly from the enclave~\circled{2}.

Note that the attestation report includes the enclave's public key~$\textit{PK}$.
V uses~$\textit{PK}$ and a nonce~$n$ to derive a symmetric encryption key~$K_U$ used only for this respective enclave and version of the model.
V encrypts the ML model using~$K_U$ and securely provisions the model to the enclave~\circled{3}.

The enclave then stores the model locally in unprotected storage~\circled{4}.
As the model can be loaded from untrusted local storage, after running the preparation phase once, steps~\circled{3} and~\circled{4} can be omitted until the vendor's model is updated.

{\bf II.\ Initialization Phase.}
Thanks to never making the decrypted model directly accessible to~U, the initialization phase can be kept simple while providing strong guarantees to~V.
V can actively manage the access of~U to the model by either sending or not sending the symmetric key~$K_U$.
In case of, e.g., an expired license,~V can stop sending~$K_U$ to the enclave, making it fail to decrypt the locally stored model.
If~V decides that~U should be allowed to use the model,~V securely sends~$K_U$~\circled{5} to the enclave and the enclave decrypts the model~\circled{6}.
As the key~$K_U$ depends on the nonce~$n$, this also prevents rollback attacks for~U's locally stored model.

{\bf III.\ Operation Phase.}
In the operation phase, the actual~ML task takes place.
U can directly and securely provide voice recordings to the enclave as~\sanctuary allows secure input from peripherals like the microphone~\circled{7} by utilizing~TrustZone features as described in~\autoref{subsec:sanctuary}.
The speech data is then processed using the model, the output can be presented to the user or made available to other applications~\circled{8}.

Once in the operation phase, the system can be queried repetitively, thereby avoiding repeated preparation and initialization costs as well as interaction with~V.
To do this, after a query is processed, the \sanctuary core can be reallocated to the commodity OS while the memory is still locked such that no device or core is able to access it.
When receiving a new query, a new \sanctuary core is allocated and the locked memory is mapped to it for performing the~ML task.

\section{Evaluation}
\label{sec:evaluation}

We demonstrate the practicality of our approach by providing a fully functional prototype implementation of~\coolname on an~ARM~HiKey~960 development board based on~TensorFlow~Lite for Microcontrollers~\cite{TensorFlowLite} and evaluating our prototype with an offline keyword recognition application.

The~ARM~HiKey~960 development board is equipped with an~ARMv8 octa-core~SoC~($4$~cores~@~\SI{2.4}{\giga\hertz}, $4$~cores~@~\SI{1.8}{\giga\hertz}) with~\SI{3}{\giga\byte} of~RAM, which closely resembles the specifications of today's mobile devices.
We use such a development board instead of an off-the-shelf device since most vendors restrict developer access to TrustZone, which prevents us from setting up \sanctuary (cf.~\autoref{subsec:sanctuary}).
As our offline keyword recognition application is just a proof of concept, following~\cite{VoiceGuard}, we do not focus on best accuracy, but study whether accuracy and runtime are affected when providing strong security guarantees.

The models are trained and evaluated on the~Speech~Command dataset~\cite{SpeechCommands} consisting of~\numprint{105000}~WAVE audio files of people saying~$30$ different words.
The recordings were post-processed to be a single word per file at a fixed~\SI{1}{\second} duration.

We follow the~TensorFlow~Lite example recipe~\cite{TensorFlowLite}:
Features are computed using a~$256$ bin fixed point~FFT across~\SI{30}{\milli\second} windows~(\SI{20}{\milli\second} shift), averaging~$6$ neighboring bins, resulting in~$43$ values per frame.
The~$49$ frames for each recording are concatenated, forming a fixed~$49 \times 43$ compressed spectrogram (\enquote{fingerprint}) per utterance.

The network architecture resembles~\cite{SainathCNN}, but is simplified to better match embedded requirements.
The \texttt{tiny\_conv} architecture feeds the audio fingerprint to a~2D convolutional layer ($8$ filters, $8 \times 10$, $x$ and $y$ stride of $2$), followed by ReLU activation and a regular layer that maps to the output labels.
During training, dropout is applied after the convolution layer.

We trained a system for a $12$-class problem: \textit{silence}, \textit{unknown}, \enquote{yes}, \enquote{no}, \enquote{up}, \enquote{down}, \enquote{left}, \enquote{right}, \enquote{on}, \enquote{off}, \enquote{stop}, \enquote{go}.
The model is first trained using TensorFlow and subsequently converted to a TensorFlow Lite and \enquote{micro} model.
The resulting compressed model is about~\SI{49}{\kilo\byte} in size.

\begin{table}[t]
\caption{Accuracy and runtime results for running the keyword recognition with and without \coolname protection.}
\label{tab:results}
	
\centering
\scalebox{1.0}{
\begin{tabular}{lrr}
\toprule
	Model                       &          Accuracy &                 Runtime \\
\midrule
	TensorFlow Lite \enquote{micro}             & \SI{75}{\percent} & \SI{379}{\milli\second} \\
	TensorFlow Lite \enquote{micro} (\coolname) & \SI{75}{\percent} & \SI{387}{\milli\second} \\
\bottomrule
\end{tabular}
}
\end{table}

We evaluated the \enquote{micro} model on a subset of the published test set comprising $10$ examples for each class, excluding the two rejection classes \enquote{silence} and \enquote{unknown}, since sensitivity for those would typically be tuned for production.

Inference was run on a~\SI{2.4}{\giga\hertz} core of the~ARM development board both with and without~\coolname protection.
Tab.~\ref{tab:results} shows the overall accuracy for the~$10$ classes, and the respective runtimes in milliseconds.
The accuracy with and without \coolname protection is \SI{75}{\percent}, confirming the correctness of the setup.
The runtimes are very close when executed with and without \coolname protection due to the fact that the hardware-enforced two-way isolation provided by \sanctuary adds no additional overhead during execution.
Since the overall duration of the test set is \SI{100}{\second}, the real-time factor is~$0.004$x.

The runtime measurements do not include the overhead for collecting the input data from the on-device microphone. 
As described in~\autoref{sec:solution}, \coolname uses the capabilities from~\sanctuary to securely connect to sensors.
Thus, only the world switch from an~SA to the secure world to request the sensor data and the switch back to the~SA introduce some overhead.
As presented in~\cite{SANCTUARY}, the switch from an~SA to the secure world takes around~\SI{0.3}{\milli\second}.
Therefore, even in the short-running speech processing use case presented in this paper, the performance overhead introduced by reading sensor data via the secure world is negligible.

Our evaluation of a keyword recognition task using spectral fingerprints and a basic~CNN lays the groundwork to port larger and recurrent architectures as well as to study training tasks.
Since our implementation has no inherent memory limitations, it also allows to securely run more complex end-to-end systems, such as the recently released TensorFlow-based dictation model by Google~\cite{GoogleAIBlog}, making it highly practical.

\section{Acknowledgments}
\label{sec:acknowledgements}

This project has received funding from the~European Research Council~(ERC) under the~European Union's Horizon~2020 research and innovation programme~(grant agreement No.~850990~PSOTI).
It was supported by the~DFG~(HWSec, project~A.1 within the~RTG~2050~\enquote{Privacy and Trust for Mobile Users}, and P3, S2, and~E4 within~CROSSING), by the~BMBF and~HMWK within~CRISP, and by the~Intel Collaborative Research Institute for Collaborative Autonomous \& Resilient Systems~(ICRI-CARS).

\bibliographystyle{IEEEtran}
\balance
\bibliography{abbrev1,main}

\end{document}